\documentclass{Interspeech}

 \interspeechcameraready

\usepackage{multirow}

\title{Enhancing Acoustic-to-Articulatory Speech Inversion by Incorporating Nasality}

\author[affiliation={1}]{Saba}{Tabatabaee}
\author[affiliation={2}]{Suzanne}{Boyce}
\author[affiliation={3}]{Liran}{Oren}
\author[affiliation={4}]{Mark}{Tiede}
\author[affiliation={1}]{Carol}{Espy-Wilson}

\affiliation{}{
Department of Electrical and Computer Engineering, University of Maryland College Park}{USA}
\affiliation{}{Department of Communication Sciences and Disorders, University of Cincinnati}{USA}
\affiliation{}{Department of Otolaryngology-Head and Neck Surgery, University of Cincinnati}{USA}
\affiliation{}{Department of Psychiatry, Yale University}{USA}

\email{ sabatb@umd.edu, boycese@ucmail.uc.edu, orenl@ucmail.uc.edu, mark.tiede@yale.edu, espy@umd.edu}


\usepackage{comment}

\begin{document}

\maketitle
\renewcommand{\thefootnote}{}%
\footnotetext{This work was supported by NSF Grant No. BCS2141413.}%
\renewcommand{\thefootnote}{\arabic{footnote}} 
\begin{abstract}
Speech is produced through the coordination of vocal tract constricting organs: lips, tongue, velum, and glottis. Previous works developed Speech Inversion (SI) systems to recover acoustic-to-articulatory mappings for lip and tongue constrictions, called oral tract variables (TVs), which were later enhanced by including source information (periodic and aperiodic energies, and F0 frequency) as proxies for glottal control. Comparison of the nasometric measures with high-speed nasopharyngoscopy showed that nasalance can serve as ground truth, and that an SI system trained with it reliably recovers velum movement patterns for American English speakers. Here, two SI training approaches are compared: baseline models that estimate oral TVs and nasalance independently, and a synergistic model that combines oral TVs and source features with nasalance. The synergistic model shows relative improvements of 5\% in oral TVs estimation and 9\% in nasalance estimation compared to the baseline models. 
\end{abstract}

\section{Introduction}
Speech articulation is a complex activity that requires finely timed coordination across articulators (lips, tongue, jaw, velum, and glottis) \cite{stevens2000acoustic}. To recover this information from speech, a Speech Inversion (SI) system was developed that maps the speech signal not onto discrete articulator positions, but rather onto synergies of activity coordinated among articulators to achieve acoustic goals known as vocal Tract Variables (TVs) \cite{browman1986towards, saltzman1989dynamical}. The TVs are defined for the lips, tongue tip, tongue body, velum and glottis (see Figure 1 and Table 1 for more details). The kinematic state of each constrictor is defined by its corresponding degree of constriction and location coordinates, recovered as time-varying trajectories by the SI system.

\begin{table}[htbp]
\centering
\scriptsize
\caption{Constrictors and corresponding vocal tract variables.}
\begin{tabular}{|l|c|}
\hline
\textbf{Constrictor} & \textbf{Vocal Tract Variable} \\ \hline
\multirow{2}{*}{Lips}        & Lip Aperture (LA) \\ 
                             & Lip Protrusion (LP) \\ \hline
\multirow{2}{*}{Tongue Tip}  & Tongue Tip Constriction Location (TTCL) \\ 
                             & Tongue Tip Constriction Degree (TTCD) \\ \hline
\multirow{2}{*}{Tongue Body} & Tongue Body Constriction Location (TBCL) \\ 
                             & Tongue Body Constriction Degree (TBCD) \\ \hline
Velum                        & Velum (VEL) \\ \hline
Glottis                      & Glottis (GLO) \\ \hline
\end{tabular}
\end{table}

\begin{figure}[htbp]
    \centering
    \includegraphics[width=0.3\textwidth, height=0.13\textheight]{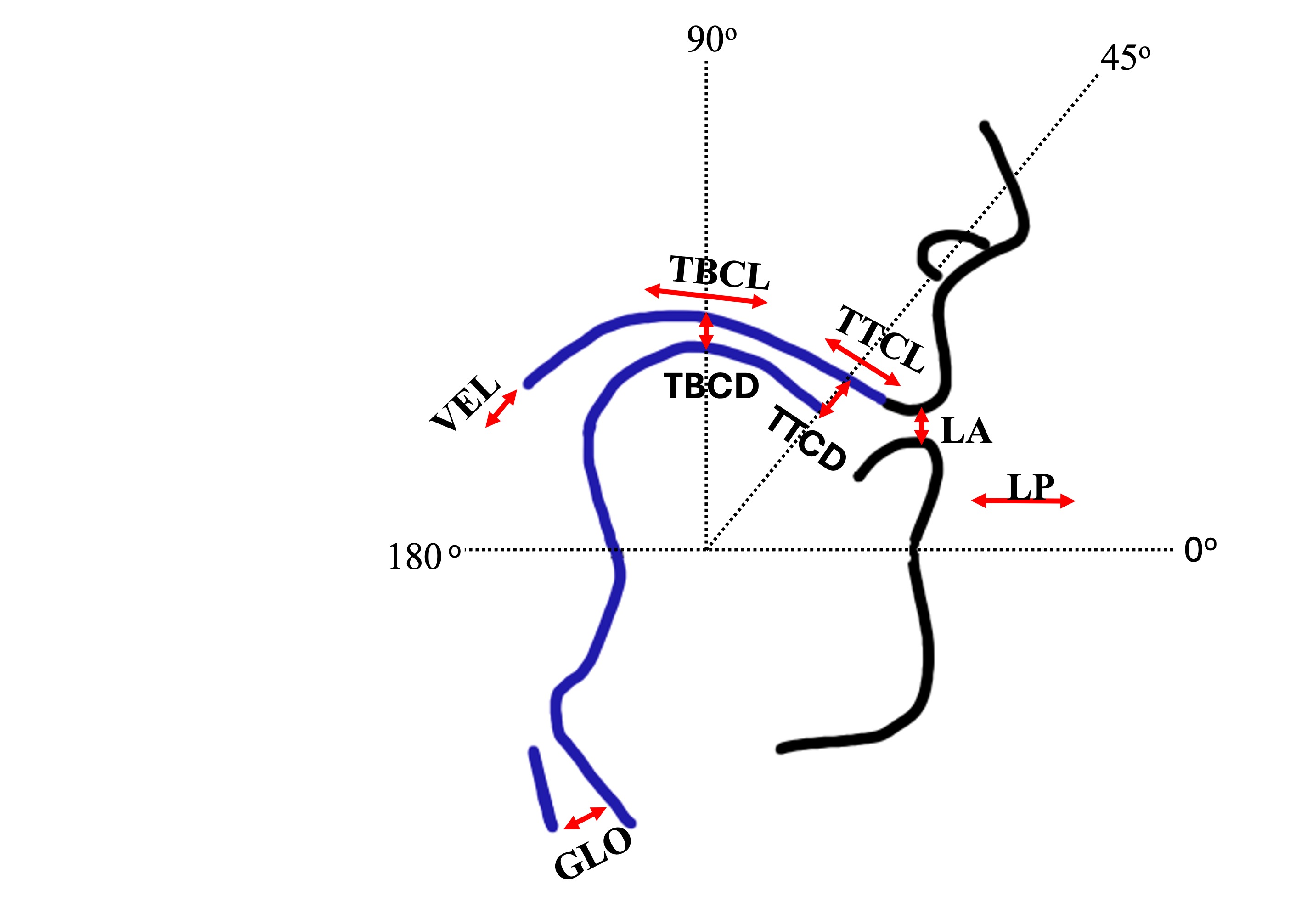}  
    \vspace{-2mm} 
    \caption{A visual representation of vocal tract variables (adapted from \cite{espy2019assessing}). }
    \label{fig:nasal_si}
    \vspace{-5mm} 
\end{figure}

The oral aspects of speech are predicted by oral TVs, which represent constrictions of the lips, tongue body, and tongue tip. Nasality is another crucial contrastive feature in speech production \cite{fry2004phonics}. Nasality is controlled by constriction of the Velopharyngeal Port (VP), which regulates airflow and acoustic coupling between the nasal and oral cavities through the coordinated movement of the velum and pharyngeal walls. Direct observations of VP movement is invasive and costly because it requires trained specialists to administer. The measure nasalance, which derives from the proportional difference in acoustic energy emitted through the nose and mouth, is an alternative for indirect measure of VP movement. A recent study \cite{siriwardena2024speaker} validated the use of nasalance as ground truth for an SI system by demonstrating strong correspondence between nasalance and VP constriction degree as measured from high-speed nasopharyngoscopy images. Therefore, in this study, we refer to the estimates of the nasalance as VP TV, using it as a proxy for VP. Previous studies have developed SI systems to estimate either oral TVs \cite{seneviratne2019multi,wu2023speaker,udupa2022streaming,fang2024performance,yan2023combining,seneviratne2018noise} or VP TV \cite{siriwardena2024speaker,feng2024wav2nas,siriwardena2023speaker} separately, without leveraging their complementary information to create a comprehensive SI system. To bridge this gap, the present study introduces a novel SI system that simultaneously estimates oral TVs plus a VP TV from speech signals. To our knowledge, this is the first study to integrate these two critical speech features into a single, unified SI framework.

Recent advancements in Self-Supervised Learning (SSL) have shifted the focus for developing SI systems towards more robust feature extraction techniques, surpassing traditional acoustic features such as Mel-frequency cepstral coefficients (MFCCs). Studies in \cite{siriwardena2024speaker, cho2023evidence, attia2024improving} demonstrate that SSL-based speech representations consistently outperform traditional acoustic features in SI tasks. For instance, one study \cite{siriwardena2024speaker} demonstrated that a BiGRNN-HuBERT system, leveraging HuBERT-Large SSL representations, outperforms a Temporal Convolutional Network (TCN) model using spectrogram inputs for VP TV estimation. The SSL-based system also exhibited superior generalization across various corpora, underscoring the robustness of SSL-derived features. Additionally, a comparison between SI systems using HuBERT-Large vs. MFCC-based methods showed that the former improves oral TVs estimation compared to the latter \cite{attia2024improving}. Further work \cite{cho2023evidence} demonstrated that WavLM-Large outperforms HuBERT-Large and other SSL models like Wav2Vec2 and TERA, as well as traditional features, in SI tasks. Building on these findings, in this work we utilize SSL representations, including HuBERT-Large and WavLM-Large, to capture richer speech features, thereby enhancing the overall performance of the SI system. 

Information in the acoustic signal is also carried by continuously varying glottal Source Features (SF). Integration of SF such as the relative amounts of non-periodic energy (aperiodicity), periodic energy (periodicity), and  fundamental frequency (F0) into a SI system was shown to improve estimation of oral TVs \cite{siriwardena2023secret}. Recent work in \cite{siriwardena2023speaker} showed that incorporation of these SF into the SI model significantly enhanced VP TV estimation. Building on these findings, this study integrates SF into the SI system to improve oral TVs and VP TV estimation.
\textbf{Our key contributions} in the current work are as follows:
\begin{itemize}
    \item Improvement of VP TV estimation by developing a Nasal-SI system with a larger training dataset, based on work in \cite{siriwardena2024speaker}.
    \item Comparative analysis of the Nasal-SI system using two SSL representation models, HuBERT-Large and WavLM-Large.
    \item Introduction of a novel comprehensive SI system capable of simultaneously estimating oral TVs, VP TV, and the SF.
    \item Comparison of multi-task learning and single-task learning approaches in developing the SI system.
    \item Providing an ablation study to analyze the effects of incorporating VP TV and the three SF in the SI system on recovery of oral TVs relative to ground truth.
\end{itemize}

\section{Dataset Description and Pre-Processing}
\subsection{Nasometry-EGG dataset}
In this study, we implemented a custom nasometry setup based on work \cite{siriwardena2024speaker}, which computes nasalance as the relative acoustic energy from separate oral and nasal microphones mounted on a plate serving to isolate the two sources. Audio signals were recorded at a 51.2 kHz sampling rate. We collected data from 24 healthy adult speakers (20 native English speakers, 3 French speakers, 1 Sinhala speaker). Electroglottography (EGG) data using electrodes placed at the thyroid prominence were collected concurrently. Participants read sections from well-known passages, including the ”Grandfather Passage” \cite{darley1975motor}, Harvard sentences \cite{rothauser1969ieee}, and other materials from \cite{krakow1989articulatory} and \cite{westbury1994speech}. The total duration was 2.1 hours, and the data were split into training (20 speakers), development (2 speakers), and test (2 speakers) sets using a speaker-independent approach.

\subsection{XRMB dataset}
The University of Wisconsin XRMB dataset \cite{westbury1994speech} contains recordings of naturally spoken isolated sentences and short passages from 32 male and 25 female participants, paired with point source trajectories of pellets attached to the tongue and lips obtained using a rasterized X-ray microbeam tracking system. After excluding mistracked data, the dataset comprises 46 native English speakers (21 males, 25 females) with approximately 4 hours of speech. Using methods described in \cite{attia2023masked}, we reconstructed much of the corrupted data, increasing the total duration to around 5.3 hours. The data were split into a training set (36 speakers) and separate development and test sets (5 speakers each, with 3 males and 2 females per set) in a speaker-independent manner. Anatomical differences between speakers cause variability in pellet positions; therefore, in this study the original X-Y coordinates are normalized for vocal tract shape by conversion into TVs using a geometric transformation from \cite{attia2024improving}. The processed XRMB dataset includes six oral TVs, sampled at 100 Hz and normalized to the range of -1 to 1. These oral TVs are: LA, LP, TBCL, TBCD, TTCL, and TTCD (see Figure 1 and Table 1 for more details).

\vspace{-3pt} 
\section{Methodology}
\subsection {Developing Nasal-SI System}
We developed the Nasal-SI system to estimate VP TV from speech signals by integrating four parameters, including the EGG envelope (EGG-env) and three SFs, which are aperiodicity (Ap), periodicity (Per), and F0, based on research in \cite{siriwardena2023speaker}. The ground truth values for the three SF were extracted using the APP detector \cite{deshmukh2005use}. Building on the approach in \cite{siriwardena2024speaker}, the Nasal-SI system was developed using a Bidirectional Gated Recurrent Neural Network (BiGRNN) based model architecture, as illustrated in Figure 2. 

\begin{figure}[htbp]
    \hfill
    \vspace{-1mm}
     \includegraphics[width=0.49\textwidth, height=0.1\textheight]{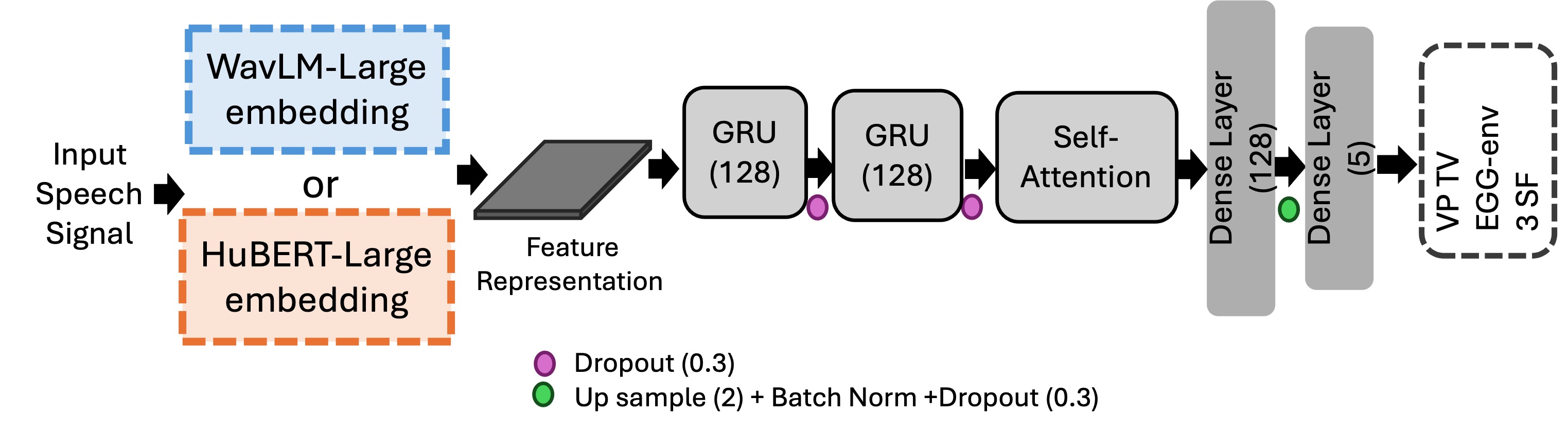}  
    \caption{Proposed model architecture for the Nasal-SI system.}
    \label{fig:nasal_si}
   
\end{figure}

The system was initially developed using HuBERT-Large \cite{hsu2021hubert} from the SpeechBrain toolkit \cite{ravanelli2021speechbrain}. Subsequently, we developed the Nasal-SI system using WavLM-Large \cite{chen2022wavlm} embedding. Both embeddings were evaluated to assess their impact on the performance of the Nasal-SI system, with the best-performing embedding model being proposed as the optimal input. The Nasal-SI model consists of two bi-directional layers of 128 Gated Recurrent Units (GRUs) with dropout layers to prevent overfitting.  After the second GRU layer, a self-attention layer is applied to enhance contextual information and capture long-range dependencies. The output is processed through a dense layer with 128 hidden units, upsampled from 50 Hz (the sampling rate of WavLM-Large and HuBERT-Large) to 100 Hz to match the target output sampling rate, and then passed through another dense layer with five hidden units to generate the outputs for VP TV, EGG-env, and three SF. The ADAM optimizer is used with a learning rate of 5e-4 and batch size of 8, selected via grid search. 
The nasalance measure used as ground truth for estimating the VP TV parameter was computed from recorded oral and nasal microphone signals by first applying a high-pass filter with a 20 Hz cutoff to remove low-frequency background noise. The acoustic energy was then calculated using the root mean square of the signals and smoothed with a 25 ms rectangular moving average filter. The nasalance measure was then computed using the nasal acoustic energy (AEnasal) and oral acoustic energy (AEoral) with equation \ref{eq:NAS}:
\begin{equation}
\text{Nasalance} = \frac{AEnasal}{AEnasal + AEoral}
\label{eq:NAS}
\end{equation}
The nasalance was downsampled to 100 Hz and normalized to the range of -1 to 1. The ground truth for EGG-env parameter was computed using the EGG signal. The EGG signal recorded at 51.2 kHz sampling rate, was first high-pass filtered at 20 Hz to remove baseline wander and noise. The EGG envelope was then extracted by computing the magnitude of the Hilbert transform. This envelope was downsampled to 100 Hz and normalized to the range of -1 to 1.
\vspace{-2.5mm} 
\subsection {Developing STL-SI and MTL-SI Systems}
The STL-SI system was developed using a Single-Task Learning (STL) approach to estimate 10 parameters, including 6 oral TVs, VP TV, and 3 SF. The MTL-SI system was developed using a Multi-Task Learning (MTL) approach, where the first task estimates the 6 oral TVs, and the second task estimates VP TV and 3 SF with a separate final layer. Both approaches were evaluated to identify the best method for developing the synergistic model that estimates these parameters simultaneously. The ground truth values for the 6 oral TVs were obtained using the method outlined in Section 2.2. For the 3 SF, the ground truth values were obtained using the APP detector \cite{deshmukh2005use}. We used the best-performing Nasal-SI system (with WavLM-Large embedding) to obtain ground truth estimates for VP TV for XRMB audios to develop the STL-SI and MTL-SI systems.

For the STL-SI and MTL-SI systems, we used the pre-trained WavLM-Large model to extract representations from input speech signals (Figure 3). The representations from 25 hidden layers of the WavLM-Large embedding were stacked, and a 2D convolutional layer computed a weighted sum of these representations, resulting in a single-layer output. This was processed through three bidirectional GRU layers (two 512-unit and one 256-unit GRU, all with a 0.3 dropout rate). This was followed by a dense layer with 128 hidden units, upsampling by a factor of 2, batch normalization, and a 0.3 dropout rate. For the final step, two approaches were explored; STL-SI used the STL approach with a dense layer with 10 units to estimate 6 oral TVs, one VP TV and 3 SF (Figure 3, Part A).  MTL-SI used the MTL approach with two final layers, one for estimating the 6 oral TVs (a dense layer with 6 units) and the other for estimating VP TV and 3 SF (a dense layer with 4 units) (Figure 3, Part B). 

\begin{figure}[htbp]
\vspace{-1pt} 
    \hfill
    \includegraphics[width=0.48\textwidth]{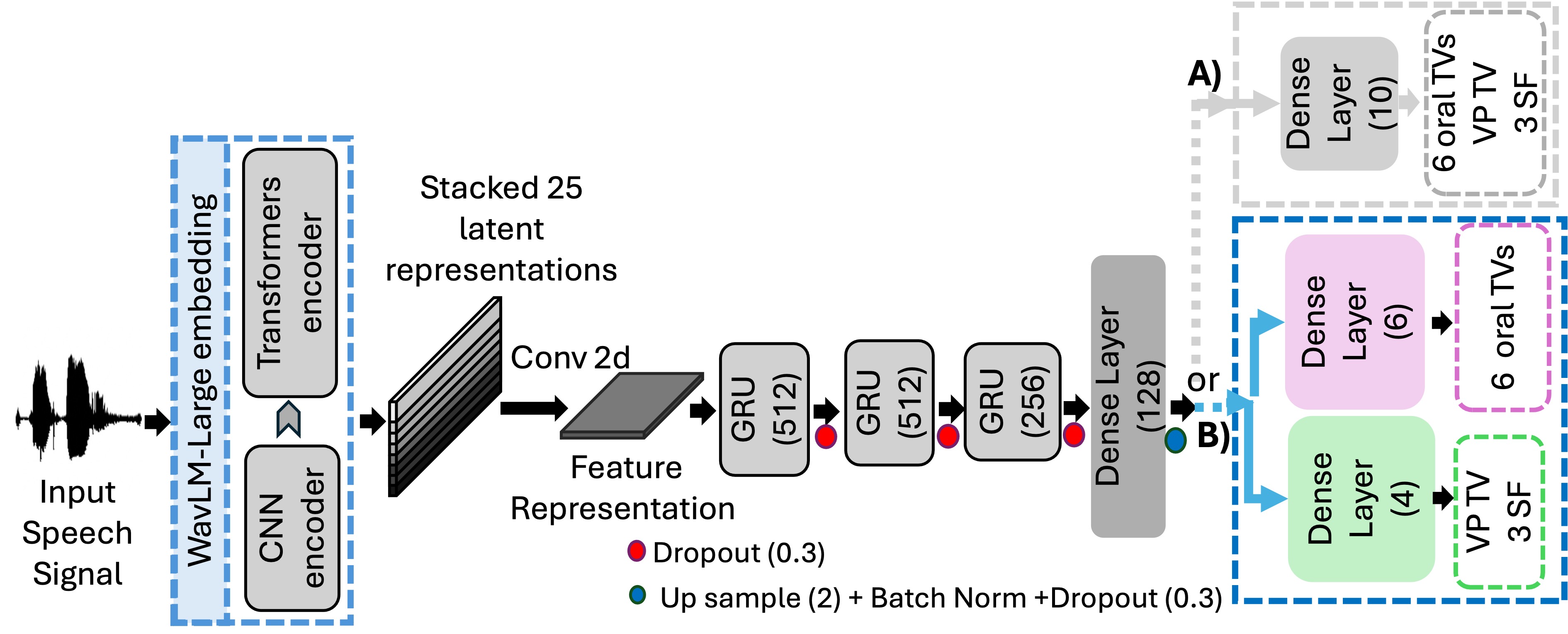}  
    \caption{Proposed model architecture for the SI systems: A) STL-SI system B) MTL-SI system.}
    \label{fig:nasal_si}
    \vspace{-1mm}
\end{figure}

STL-SI and MTL-SI were trained using the Adam optimizer with a learning rate of 5e-4 and a batch size of 8, both selected via grid search. Early stopping with a patience of 8 epochs was used to prevent overfitting. The loss function to develop STL-SI and MTL-SI is shown in Equation~\ref{eq:loss}, which combines Pearson Correlation (PC) and Root Mean Square Error (RMSE), with $\alpha$ empirically set to 0.8 for optimal performance:
\begin{equation}
\text{Loss} = \alpha (1 - \text{PC}) + (1 - \alpha) \cdot \text{RMSE}
\label{eq:loss}
\end{equation}
\section{Results and Discussion}
\subsection{Results for Nasal-SI System }
To evaluate the Nasal-SI system, we used Pearson Product-Moment Correlation (PPMC) to measure the similarity between the ground truth and estimated parameters. Building on the approach in \cite{siriwardena2024speaker}, the training set was expanded by incorporating recordings from four additional speakers, to improve the VP TV estimation. The performance of the Nasal-SI system was compared with the BiGRNN-HuBERT model from \cite{siriwardena2024speaker}, which served as the baseline, with both models evaluated on the same test set for consistency. In \cite{siriwardena2024speaker}, PPMC results for the BiGRNN-Hubert model were reported after segmenting the test set audio into fixed 2-second duration intervals. Therefore, we evaluated the Nasal-SI system under two scenarios: first, the audio in the test set was segmented into 2-second durations, and the PPMC score was computed on these intervals for the Nasal-SI system. These results were then compared to those of the BiGRNN-Hubert model. Second, the test set audio was used in its unsegmented form, with the PPMC score calculated for complete utterances. We also obtained PPMC results for the BiGRNN-HuBERT model under this scenario and compared them to those of the Nasal-SI system. 
\begin{table}[htbp]
\caption{PPMC scores of the Nasal-SI models on the Nasometry-EGG test set. H: HuBERT-Large, W:WavLM-Large.}
\scriptsize
\centering
\label{speaker_verification_models}
\resizebox{\columnwidth}{!}{
    \begin{tabular}{|l|c|c|c|c|c|c|c|c|}
        \hline
        Model & Embed.& Seg.& VP & EGG-env & Per & Ap & F0 \\
        \hline
    
         \cite{siriwardena2024speaker}  &H & - &0.8757 & 0.8258 & 0.7309 & 0.5307 & 0.7644 \\
        
        \hline 
          Nasal-SI &H & - & 0.8904 & 0.8331 & 0.7320 & 0.5348 & 0.7669 \\
        \hline
         Nasal-SI &W & - & 0.9152 & 0.8403 &  0.6683 &  0.5392 &  0.6957 \\
        \hline
         \noalign{\vskip 0.1cm}  
         \hline
        \cite{siriwardena2024speaker}   &H & 2& 0.8115  & 0.8330  & 0.8373 & 0.8542  & 0.8562  \\
        \hline
         Nasal-SI & H &2&0.8533 & 0.8145 &0.7648& 0.5687& 0.7214\\
        \hline
         Nasal-SI &W & 2&0.8663 & 0.8425 & 0.7673 & 0.6111 & 0.7209 \\
        \hline       
    \end{tabular}
    }
\end{table}

\begin{table*}[htbp]
\caption{PPMC score for estimated parameters of SI models on XRMB test set.}
\scriptsize
\centering 
\label{speaker_verification_models}

    \begin{tabular}{|l|c|c|c|c|c|c|c|c|c|c|c|c|c|}
        \hline
        Model &Embedding &  VP & LA & LP & TBCL & TBCD & TTCL & TTCD  & Per & Ap& F0 & AVG. 
 oral TVs   \\
        \hline   
            \cite{attia2024improving} & HuBERT-Large& - & 0.8902& 0.7142 & 0.7361& 0.8180 & 0.8032 & 0.9229 & - & - & - & 0.8141   \\
        \hline
     STL-SI  & WavLM-Large &0.9420 & 0.9026 & 0.7376 &0.7591 &   0.8509 & 0.8307 &
 0.9403 & 0.9326 &  0.8759 &  0.7551& 0.8372\\
     \hline
     MTL-SI  & WavLM-Large &0.9462 & 0.9104 & 0.7594 & 0.7981 & 0.8626 &  0.8360 & 0.9478 & 0.9403 & 0.8815 & 0.7470 & 0.8524\\
     \hline
    \end{tabular}
    \vspace{-3mm} 
\end{table*}

As shown in Table 2, the Nasal-SI system with the WavLM-Large model outperforms alternative configurations, achieving the highest PPMC score across both experimental scenarios. Specifically, the relative improvements compared to the baseline BiGRNN-HuBERT are 6.75\% and 4.51\% for segmented and unsegmented audio conditions, respectively. Moreover, the results show that using WavLM-Large embeddings improves performance in VP TV estimation compared to Hubert-Large embeddings when developing the Nasal-SI system. This finding aligns with work in \cite{chen2022wavlm}, which highlighted the superior performance of WavLM-Large embeddings over Hubert-Large embeddings in ASR and other tasks. In this study, we extend this conclusion by demonstrating that WavLM-Large significantly boosts performance in SI systems, particularly for VP TV estimation, a domain not explored in \cite{chen2022wavlm}.
\subsection{Results for the STL-SI and MTL-SI Systems } 
\subsubsection{Oral TVs Estimation in STL-SI and MTL-SI Models}
The SI system in \cite{attia2024improving}, which employs HuBERT-Large SSL representations, serves as the baseline for comparison with the STL-SI and MTL-SI systems, since it applies the same geometric transformations to generate ground truth values for the oral TVs as those used in the STL-SI and MTL-SI systems. Additionally, to ensure a fair comparison, we used the same train, test, and development splits. As shown in Table 3, developing the SI system with the MTL approach resulted in the best performance for oral TVs parameter estimation, outperforming both the STL-SI system and the baseline model. Specifically, MTL-SI achieved a 4.70\% relative improvement in the average PPMC score for oral TVs compared to the baseline model.

Table 4 presents an ablation study of the MTL-SI system, the top-performing model, to identify the effect of SF and VP TV on oral TVs estimation. In the first row, we excluded the VP TV and 3SF parameters, estimating only the 6 oral TVs, which created a single-task setup similar to the baseline model. Compared to the baseline, the average PPMC score for 6 oral TVs increased from 0.8141 to 0.8411, highlighting the effectiveness of the proposed SI system design. In the subsequent steps, we estimated either the VP TV or 3SF along with the 6 oral TVs and trained the MTL-SI model in each case. In both scenarios, the PPMC scores improved, demonstrating that incorporating additional speech information, whether 3SF or nasalance, enhances the estimation of oral TVs. Finally, the best performance in oral TVs estimation was achieved when 6 oral TVs were estimated together with VP TV and 3SF. This demonstrates that integrating complementary phonetic information into the MTL framework enhances the accuracy of oral TVs estimation, which is consistent with recent works \cite{wu2023speaker, siriwardena2022acoustic}.
 \vspace{-2mm} 
\begin{table}[htbp]
\caption{Ablation study of the MTL-SI system: PPMC scores for different parameter exclusions on the XRMB test Set.}
\vspace{-1mm} 
\scriptsize
\centering 
\label{ablation}
    \begin{tabular}{|l@{\hskip 2pt}|c@{\hskip 4pt}|c@{\hskip 4pt}|c@{\hskip 3pt}|c@{\hskip 3pt}|c@{\hskip 3pt}|}
        \hline
        Excluded Param & VP & AVG. oral TVs  &Per& Ap& F0  \\
        \hline
     VP , 3 SF  &- &0.8411& - & - & - \\
     \hline
     3 SF& 0.9503& 0.8485 & - & - & - \\
     \hline
        VP    &- &0.8489 &
0.9437 &0.8894 &  0.7678 \\
             \hline
-   &0.9462 &0.8524 &
0.9403 & 0.8815 & 0.7470\\
             \hline             
    \end{tabular}
    
\end{table}
 
\subsubsection{VP TV Estimation in the STL-SI and MTL-SI Models}
We evaluated the STL-SI and MTL-SI systems for VP TV estimation using the Nasometry-EGG test set, with ground truth derived from recorded oral and nasal signals. We evaluated the STL-SI and MTL-SI systems in two scenarios: using 2-second audio segments and the original unsegmented audios from the test set. The results of the best-performing Nasal-SI system (with WavLM-Large) for both scenarios are also presented for comparison in Table 5. The results in Table 5 indicate that, for each of the two scenarios, both the STL-SI and the MTL-SI systems outperform the Nasal-SI system. The MTL-SI system outperforms STL-SI in VP TV estimation, highlighting that MTL not only improves oral TVs estimation (as shown in Table 3) but also enhances VP TV estimation. The MTL-SI model outperformed the BiGRNN-HuBERT \cite{siriwardena2024speaker} (used as the baseline model in Table 2) by relative improvements of 5.82\% in the unsegmented audio scenario and 9.08\% in the segmented audio scenario. Overall, the MTL-SI system achieved high PPMC scores for both oral TVs and VP TV. This suggests that the MTL-SI system, by consolidating the estimation of VP TV and oral TVs into a single model, is more powerful compared to SI systems that estimate them separately.

\begin{table}[htbp]
\caption{PPMC scores for VP TV estimation of MTL-SI and STL-SI models on the Nasometry-EGG test set. }
\vspace{-1mm}
\scriptsize
\centering 
\label{speaker_verification_models}
    \begin{tabular}{|l|c|c|c|c|c|c|c|c|}
        \hline
        Model & Embedding& Segment& VP\\
        \hline
       Nasal-SI &WavLM-Large & - & 0.9152  \\
        \hline
              STL-SI  &WavLM-Large & - & 0.9193  \\
    
        \hline
             MTL-SI &WavLM-Large & - &0.9267  \\
         \hline

         \noalign{\vskip 0.1cm}  
         \hline

          Nasal-SI &WavLM-Large & 2 Second&0.8663  \\
        \hline       
         STL-SI  &WavLM-Large  &  2 Second&0.8802 \\
        \hline 
         MTL-SI  &WavLM-Large  &  2 Second &0.8852  \\
        \hline
    \end{tabular}
    \vspace{-2mm} 
\end{table}

\subsubsection{Cross-corpus evaluation of the proposed MTL-SI system }
To further test generalization, we compared the outputs from the MTL-SI system for an utterance “Say packed memos” taken from a different dataset for which nasometry and Electromagnetic Articulography (EMA) were used to simultaneously collect audio and the X-Y movements of point-source sensors placed on the lips, Tongue Tip (TT) and Tongue Body (TB). Using the oral and nasal signals, we obtained the nasalance as defined in equation 1. Figure 4 shows that the MTL-SI system accurately estimates the constriction patterns for each consonant, closely matching the ground truth. Furthermore, for the two instances of /m/ in the ”memos”, the VP TV estimated by the MTL-SI system exhibits two peaks, consistent with the nasalance and co-located with the lip closure for the two /m/. These results suggest that the MTL-SI system can generalize effectively to a corpus that was not encountered during training.
\begin{figure}[htbp]
    \hfill
    \includegraphics[width=0.45\textwidth, height=0.25\textheight]{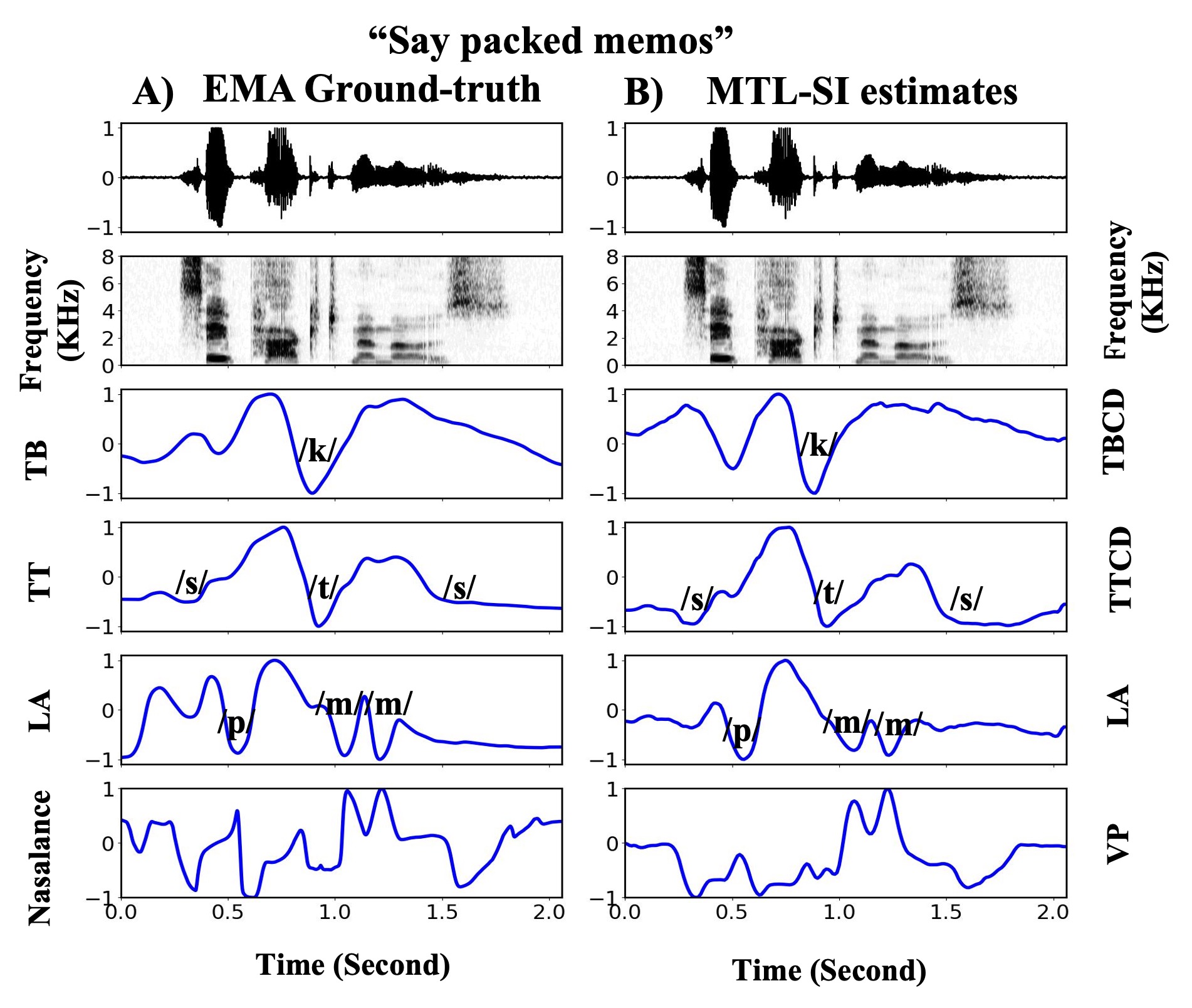}  \vspace{-2mm} 
    \caption{Waveforms, spectrograms, followed by a comparison of the ground-truth EMA and nasalance with the corresponding estimates from the MTL-SI system for the utterance “Say packed memos”. The second row shows the spectrogram, visualizing the signal's frequency content over time with color intensity representing frequency strength. The primary constriction for each consonant is labeled. Note that the peaks in VP match the labial constrictions for the two /m/ in “memos”.  }
    \label{fig:perfectmemory}
    \vspace{-3mm}
\end{figure}
\vspace{-2mm}
\section{Conclusions And Future Work}
In this paper, a novel SI system was proposed that simultaneously estimates VP TV, along with oral TVs and three source features.
The results demonstrated that the proposed synergistic SI model improves the estimation of VP TV and oral TVs, underscoring the complementary nature of these characteristics. Additionally, the results highlighted the effectiveness of the multi-task learning framework and the use of WavLM-Large self-supervised representations to boost the performance of the SI system. In particular, the MTL-SI model outperformed other SI systems that estimate VP TV and oral TVs separately, showcasing the advantages of integrating multiple tasks into a single model. The MTL-SI system, by estimating multiple speech parameters within a single framework, has clinical potential for effective monitoring and intervention, particularly for craniofacial disorders. Future work will focus on its application in the clinical diagnosis of velopharyngeal port dysfunction.

\bibliographystyle{IEEEtran}
\bibliography{mybib}

\end{document}